# Comparison of Quantum Simulators for Variational Quantum Search: A Benchmark Study

Mohammadreza Soltaninia, Junpeng Zhan
Department of Electrical Engineering, Alfred University, Alfred, NY 14802, USA

*Abstract*—Simulating quantum circuits using classical computers can accelerate the development and validation of quantum algorithms. Our newly developed algorithm, variational quantum search (VQS), has shown an exponential advantage over Grover's algorithm in the range from 5 to 26 qubits, in terms of circuit depth, for searching unstructured databases. We need to further validate the VQS for more than 26 qubits. Numerous simulators have been developed. However, it is not clear which simulator is most suitable for executing VQS with many qubits. To solve this issue, we implement a typical quantum circuit used in VQS on eight mainstream simulators. Results show that the time and memory required by most simulators increase exponentially with the number of qubits and that Pennylane with GPU and Qulacs are the most suitable simulators for executing VQS efficiently. Our results aid researchers in selecting suitable quantum simulators without the need for exhaustive implementation, and we have made our codes available for community contributions.

## I. Introduction

Quantum computing (QC) shows great promise for revolutionizing various fields by offering enhanced computational capabilities compared to classical computers. However, the accessibility of *real quantum computers* is limited, often resulting in long wait times. Moreover, the finite entanglement time of qubits imposes constraints on the achievable circuit depth.

As a result, *simulators* play a crucial role in accelerating the development and validation of quantum algorithms, providing a valuable platform for algorithm exploration and noise simulation, and insights into the behavior of real quantum systems. However, choosing the right simulator can be challenging, especially for beginners, as there are multiple options available, each with its features and limitations. This exploration process can be time-consuming and may lead to the selection of a simulator that is not well-suited to the specific requirements of the research.

In this paper, we focus on the *Variational Quantum Search* (VQS) algorithm [1], which has demonstrated exponential advantage over Grover's search algorithm and has been successfully verified up to 26 input qubits. However, an important question remains: can this exponential advantage be scaled up to around 50 qubits [2]? To address this question, we investigate various simulators to determine if there is one that can simulate the VQS with many qubits.

Our first contribution is a comprehensive comparison of commonly used simulators, providing *recommendations* on which simulators to choose based on their respective strengths and limitations. We assess computational time and memory usage to guide researchers, particularly beginners, in selecting the most suitable simulator for their simulations.

Furthermore, we explore the performance of different simulators and find that all of them encounter challenges related to exponentially increasing time or memory requirements as the number of qubits grows.

## II. Problem and Simulator Description

In this study, we execute a typical quantum circuit used in VQS on different simulators, i.e., calculate the expectation value of the observable $Z_1$, denoted as $\langle Z_1 \rangle$, as shown in Fig. 1a of Ref [1]. Note that this paper uses type-II Ansatz with three layers for the VQS. Calculating the $\langle Z_1 \rangle$ is the most time and memory-consuming part of executing VQS on a classical computer. The maximum memory required for simulating VQS is exactly the memory needed for calculating $\langle Z_1 \rangle$.

We have executed the quantum circuit on eight common simulators which are briefly described below.

*Qiskit*: An open-source framework by IBM for QC research, offering a user-friendly interface, versatile functionality, and support for both simulation and execution on real quantum hardware [3].

*Pennylane*: An open-source library that combines classical machine learning with QC, enabling the construction and training of quantum neural networks. It integrates with popular frameworks and supports CPU and GPU computation [4].

*TensorCircuit*: A Python-based QC framework emphasizing speed and flexibility. It provides efficient simulations of quantum circuits and seamless integration with machine learning frameworks like TensorFlow and JAX [5].

*Qulacs*: A powerful and versatile QC framework with high-performance simulation capabilities. It supports both CPU and GPU computing, offering a user-friendly interface and a variety of quantum gates and operations [6].

*ProjectQ*: An open-source software framework with compilation and simulation capabilities. It allows running quantum programs on IBM Quantum Experience chip, AWS Bracket, Azure Quantum, or IonQ service provided devices [7].

*Cirq*: An open-source Python framework for writing, manipulating, and optimizing quantum circuits. It focuses on near-term quantum algorithms, offering fine-grained control over circuits and compatibility with quantum computers and simulators [8].

## III. Results

We calculated the exact expectation value described above for the VQS on various simulators using the NCSA Delta high performance computer (A100x8). We adopt the most appropriate configuration for implementing VQS in each simulator. Fig. 1 displays the computational time and memory usage for different qubit numbers.

We mainly focus on simulating larger systems, particularly those with more than 26 qubits. As shown in Fig. 1, we compare the performance of different simulators in terms of resource consumption. The results demonstrate that Pennylane using GPU exhibits the lowest time consumption, followed by Qulacs (see Fig. 1a). In terms of memory usage, Qulacs requires the least amount, followed by Pennylane using GPU (see Fig. 1b).

Figure 1 shows that the time consumption of each simulator increases exponentially as the number of qubits increases. Specifically, adding one qubit roughly doubles the time consumption. On the other hand, except for Qulacs with the "CasualConeSimulator" backend, the memory consumption of all simulators increases exponentially with the number of qubits. Notably, the Qulacs simulator consistently requires only 0.02 MiB across all cases involving 24-30 qubits, primarily because it selectively extracts only the necessary gates linked to a specific observable by reversing the circuit traversal [6].

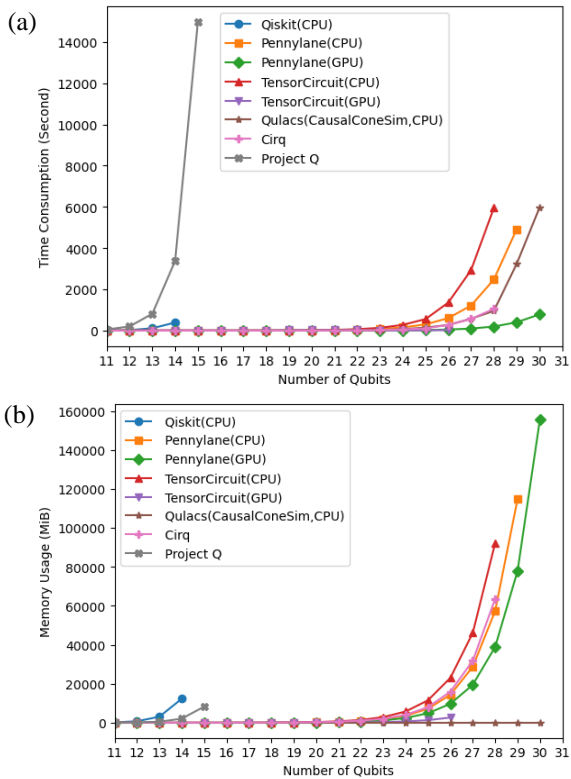

Figure 1: Time (top panel) and memory (bottom panel) consumed by different simulators to obtain the expectation value of observable $\langle Z_1 \rangle$ in the VQS for different numbers of qubits. Note: *Pennylane (CPU)*, *Pennylane (GPU)*, and *TensorCircuit (CPU)* reached their memory limit for more than 29, 30, and 28 qubits, respectively. *Qiskit (CPU)* and *Cirq* encounter errors when calculating the exact expectation value for more than 15 and 28 qubits, respectively. *Qulacs* and *Project Q* encounter time limits when calculating the exact expectation value for more than 30 and 16 qubits, respectively.

Another notable observation is that TensorCircuit with GPU has the best performance in terms of time consumption (refer to table results in [10]) while ranking second in memory consumption (behind Qulacs) for up to 26 qubits. Although TensorCircuit with GPU encounters tensor limitations for circuits larger than 26 qubits, its potential as a simulator for superior simulation of larger quantum circuits is evident.

Based on our findings, we recommend using Qulacs for optimal memory efficiency and Pennylane with GPU for optimal time efficiency.

## IV. CONCLUSION AND FUTURE WORK

Our benchmarking of quantum simulators for VQS provides valuable insights into their scalability and efficiency. Most simulators exhibit exponential growth in time and memory consumption with the number of qubits, except for Qulacs. This necessitates the exploration of alternative techniques, such as circuit cutting or Matrix Product State, to enable simulations of VQS on a larger scale. Pennylane with GPU is the optimal choice for time-constrained scenarios, while Qulacs excels in minimizing memory usage.

The results presented above focus on calculating the exact expectation value without considering noise and sampling effects. The exact simulation serves the purpose of validating the correctness of quantum algorithms, rather than assessing their performance on real quantum hardware. We plan to explore sampling-based simulations, where noise and sampling effects are considered, in future research.

Our findings have revealed that TensorCircuit with GPU is highly efficient in time and memory usage for up to 26 qubits. However, GPU memory allocation becomes challenging for more qubits. To overcome this, we plan to utilize multiple GPUs and leverage CUDA-specific instructions [9] for effective GPU memory management.

We have shared our codes on GitHub [10], enabling community contributions to expand the comparison of quantum simulators. This establishes our work as a foundation for an ongoing project to benchmark diverse simulators.

## ACKNOWLEDGMENTS

We acknowledge the support from the NSF ERI program, under award number 2138702. This work used the Delta system at the National Center for Supercomputing Applications through ACCESS allocations CIS220136 and ELEC220008.